# Nucleation and propagation of thermomagnetic avalanches in thin-film superconductors (Review Article)

J. I. Vestgården[a)]

*Department of Physics, University of Oslo, P.O. Box 1048 Blindern, Oslo 0316, Norway and Norwegian Defense Research Establishment (FFI), Kjeller, Norway*

T. H. Johansen[b)]

*Department of Physics, University of Oslo, P.O. Box 1048 Blindern, Oslo 0316, Norway and Institute for Superconducting and Electronic Materials, University of Wollongong Northfields Ave., Wollongong, NSW 2522, Australia*

Y. M. Galperin[c)]

*Department of Physics, University of Oslo, P.O. Box 1048 Blindern, Oslo 0316, Norway and Ioffe Physical Technical Institute, 26 Polytekhnicheskaya, St Petersburg 194021, Russia*

Stability of the vortex matter—magnetic flux lines penetrating into the material—in type-II superconductor films is crucially important for their application. If some vortices get detached from pinning centres, the energy dissipated by their motion will facilitate further depinning, and may trigger an electromagnetic breakdown. In this paper, we review recent theoretical and experimental results on development of the above mentioned thermomagnetic instability. Starting from linear stability analysis for the initial critical-state flux distribution we then discuss a numerical procedure allowing to analyze developed flux avalanches. As an example of this approach we consider ultra-fast dendritic flux avalanches in thin superconducting disks. At the initial stage the flux front corresponding to the dendrite's trunk moves with velocity up to 100 km/s. At later stage the almost constant velocity leads to a specific propagation regime similar to ray optics. We discuss this regime observed in superconducting films coated by normal strips. Finally, we discuss dramatic enhancement of the anisotropy of the flux patterns due to specific dynamics. In this way we demonstrate that the combination of the linear stability analysis with the numerical approach provides an efficient framework for understanding the ultra-fast coupled nonlocal dynamics of electromagnetic fields and dissipation in superconductor films. *Published by AIP Publishing.* https://doi.org/10.1063/1.5037549

*In memory of Aleksei Alekseevich Abrikosov*

## 1. Introduction

A very important feature of superconductors is the Meissner and Ochsenfeld effect—expulsion of weak external magnetic fields, $H$, from their interior. Therefore, a superconductor in weak external magnetic fields behaves as a perfect diamagnet. In type-II superconductors, the perfect diamagnetism exists for applied fields below a lower critical field, $H_{c1}$, and there is a broad domain of magnetic fields, $H_{c1} \leq H \leq H_{c2}$, where the field penetrates the sample in the form of quantized flux lines—Abrikosov vortices.[1] An isolated vortex consists of a core where the superconducting order parameter is suppressed, while the magnetic field reaches a local maximum. The radius of the core is of the order of the coherence length, $\xi$. Outside the core the magnetic field decays exponentially over a distance of the magnetic penetration depth, $\lambda_L$, where also electrical current circulates. Each vortex carries one flux quantum $\Phi_0 = h/2e \approx 2.07 \times 10^{-15}$ Wb.

Parallel flux lines repel each other, an interaction that can be understood by applying Ampère's law to the circular currents. The repulsion leads to formation of a flux line lattice. In a perfect sample this so-called Abrikosov lattice is regular. A number of phases and dynamic effects in the flux line lattice was reviewed in Refs. 2 and 3. Above the upper critical field, $H_{c2}$, the bulk superconductivity seizes to exist.

The vortices interact with an electrical current via the Lorentz force per unit length

$$\mathbf{f} = \Phi_0 [\mathbf{j} \times \mathbf{n}], \qquad (1)$$

where $\mathbf{j}$ is the current density and $\mathbf{n}$ is the unit vector along the flux line. Since vortex motion implies displacement of the vortex cores containing quasiparticles, the motion is accompanied with dissipation. At small velocities the dissipation is proportional to the velocity, therefore the dissipation can be described by an effective viscosity. The velocity is determined by the balance between the Lorentz force and the viscous force. Therefore, a free vortex lattice would move as a whole with a constant velocity, and result in a finite resistance of the sample. Such a vortex lattice is said to be in the flux flow state.

However, in real superconductors the flux lines interact with material defects that will act as pinning centers and thus hamper the flux line motion. Pinning barriers often arise from rather inevitable structural irregularities such as vacancies, dislocations, grain boundaries, etc. In addition, there exists a rich zoo of artificially introduced pinning centers. Among them are magnetic inclusions, phases of weaker or no superconductivity, lithographically patterned "antidots," magnetic dots, etc. According to the particular nature and dimensionality of the defects the pinning potential has different spatial extent and different dependence on magnetic field and temperature, see Ref. 4 for a review.

When a superconductor is exposed to an increasing external magnetic field, or self field of a transport current, vortices form at the edges and then propagate inwards. The presence of pinning leads to formation of an inhomogeneous distribution of the magnetic flux. According to the critical state model[5] the stationary distribution can be found from Ampère's law with the condition that the current density at each point is equal to its local critical value, $j_c(\mathbf{B},T)$, i.e.,

$$\nabla \times \mathbf{B} = \mu_0 \mathbf{j}, \quad |\mathbf{j}| = j_c(\mathbf{B},T), \tag{2}$$

where $\mathbf{B}$ is the magnetic induction.

The case where $j_c$ is independent of $\mathbf{B}$ is called the Bean model.[5] The energy loss for $j < j_c$ is typically very low. Therefore, $j_c$ is a key measure of the performance of superconductors. Microscopic evaluation of the critical current density is an extremely difficult task since it requires direct summation of vortex-vortex interactions and all elementary pinning forces. Thus, the critical state model with phenomenological $j_c(\mathbf{B},T)$ has become a major paradigm in the studies of electromagnetic properties of type-II superconductors.

The critical state model is valid also in thin films, but when doing calculations one must include the film self-field. As a result, exact calculations are possible only for a few geometries, such as long strips,[6] rectangles[7] and circular disks.[8,9] A consequence of the self-field is the flow of shielding currents with $j < j_c$ in the parts of the sample where $B_z = 0$. Moreover, in films the profiles of $B_z$ are much different from in bulks, as $B_z$ in films has a non-trivial shape showing large field amplification along the edge. Such field enhancement is seen in Fig. 1 (upper panel), presenting a magneto-optical image of a square film of $YBa_2Cu_3O_x$ where flux has penetrated equally from each edge. The penetration forms a tongue-like pattern from each edge, consistent with the critical-state model.[7] The black central area shows the flux-free region.

An important feature of the critical state is that it is metastable, i.e., an increase in the external magnetic field may lead to collapse by a sudden large-scale redistribution of the flux. Experimentally, such dramatic events can be observed as abrupt drops in the magnetization, so-called *flux jumps*. They are commonly ascribed to a thermomagnetic instability where the local heat release associated with vortex motion reduces the pinning, which in turn facilitates further vortex motion. With this positive feedback, a small perturbation can quickly evolve into a macroscopic avalanche.

In thin films such avalanches form fingering and branching structures, see, e.g., Refs. 10–23. An example is presented in Fig. 1 (lower panel), where the image shows a 400 nm thick film of $MgB_2$ initially zero-field-cooled to 9.9 K. Then, while slowly ramping the perpendicular applied magnetic field, the seen dendritic flux structure abruptly appeared at $\mu_0 H = 17$ mT. Redoing the experiment, the qualitative behavior repeated, but the dendritic pattern was always different.

Another key experiment was reported by Baziljevich et al.,[24] who investigated avalanche activity in films of $YBa_2Cu_3O_x$ deposited on a strontium titanate substrate. When a 150 nm thick film was exposed to a perpendicular field ramped at the rate of 3000 T/s, a highly dramatic avalanche event occurred. Examining the film afterwards using AFM, it was found that the advancing dendrites had caused the local temperature to rise so high that the material decomposed, thus providing a clear manifestation of the thermomagnetic nature of the phenomenon. In the following, we present more experimental results supplemented by explanations based on analytical theory, as well as numerical simulations.

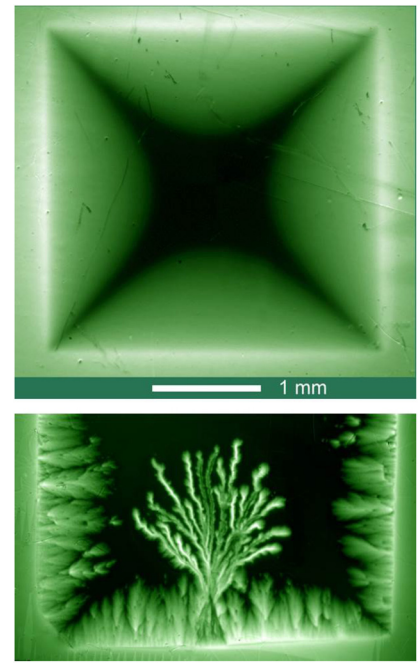

Fig. 1. Upper panel: Magneto-optical image of the magnetic flux distribution in a square film of $YBa_2Cu_3O_x$ exposed to a perpendicular magnetic field of 20 mT. Lower panel: Flux distribution in a $MgB_2$ film after a dendritic avalanche occurred from the lower edge. The image brightness represents perpendicular component of the magnetic induction, $B_z$.

The paper is organized as follows. In Sec. 2 we briefly describe the experimental method of magneto-optical imaging (MOI), while Sec. 3 presents the characteristic features of the observed avalanche behavior. Then, Sec. 4 gives a linear stability analysis of superconducting films, which for generality are coated with a layer of normal metal. In Sec. 5 the methods for numerical modeling are presented and with Secs. 6–8 presenting and discussing different examples of flux propagation. In Sec. 6 we report on ultra-fast propagation of dendrites in superconducting disks while Sec. 7 is aimed at specific propagation of the flux avalanches resembling ray optics. In Sec. 8 we discuss observed dramatic anisotropy of the flux avalanches and provide relevant theoretical explanation. We conclude the reported results in Sec. 9.

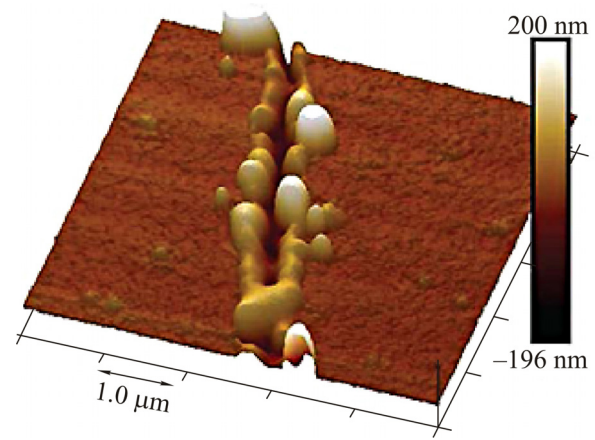

Fig. 2. Height profile plot obtained by AFM scan of a $YBa_2Cu_3O_x$ film after being exposed to a rapidly increasing perpendicular applied magnetic field. From Ref. 24.

## 2. Experimental

Experimental methods employed to investigate the avalanches in the vortex matter can be subdivided in two groups: integral and spatially resolved.

Integral methods include many types of magnetometry: inductive coils, vibrating sample magnetometry and SQUID magnetometry.[25] These measurements are sensitive to global redistributions of the flux and current flow, and in particular, they detect the change in the total magnetic moment caused by an avalanche taking place anywhere in the sample.

A disadvantage of the integral methods is a lack of detailed information about the avalanche events, e.g., their location in the sample, their morphology, etc. Moreover, the relatively low sampling rate makes it difficult to separate events occurring within short time intervals, and impossible in the case of simultaneous avalanches. It can also be difficult to discriminate between small jumps and instrument noise. These problems are partly solved in spatially resolved magnetometry; an overview of available methods can be found in Ref. 26. Recently, an ultrafast spatially resolved SQUID magnetometer was developed[27] and applied to investigation of flux avalanches in their initial stage when the vortex motion is very fast.[28]

Among the space-resolved methods, one of the most powerful is magneto-optical imaging (MOI), which combines high magneto-spatial resolution and short acquisition time. Figure 3 illustrates the principal experimental scheme used for most MOI studies of flux dynamics in superconductors, and is based on polarized light microscopy.[25,29]

As sensor one uses a layer of Faraday-rotating material placed in close proximity to the sample under investigation.[30] The MOI results reported in this paper were obtained using the large Faraday rotation in ferrite garnet films (FGFs) of composition $(Lu,Bi)_3(Fe,Ga)_5O_{12}$. These films were grown as a few micron thick epitaxial layer on optically transparent gadolinium gallium garnet substrates, where the FGFs become spontaneously in-plane magnetized.[31,32]

The presence of perpendicular flux in the sample under investigaton will in the adjacent FGF locally tilt the magnetization vector out-of-plane creating a distribution of Faraday rotation angles in the polarized light passing through the indicator chip. After reflection by a mirror deposited on the FGF, or from the sample itself if its surface is well reflecting, the Faraday rotation is doubled. When then passing a crossed analyzer an image is formed where the brightness is a direct measure of the magnetic flux distribution in the plane of the sample surface. The image is recorded by a CCD camera.

The sensitivity of the FGFs is characterized by the Verdet constant, which for the films used in the works reviewed here are (2–8) $\times$ $10^{-2}$ deg/mT per micron thickness. Their dynamic range is limited upwards to approximately 100 mT, when the FGF reaches saturation by becoming magnetized fully out-of-plane.

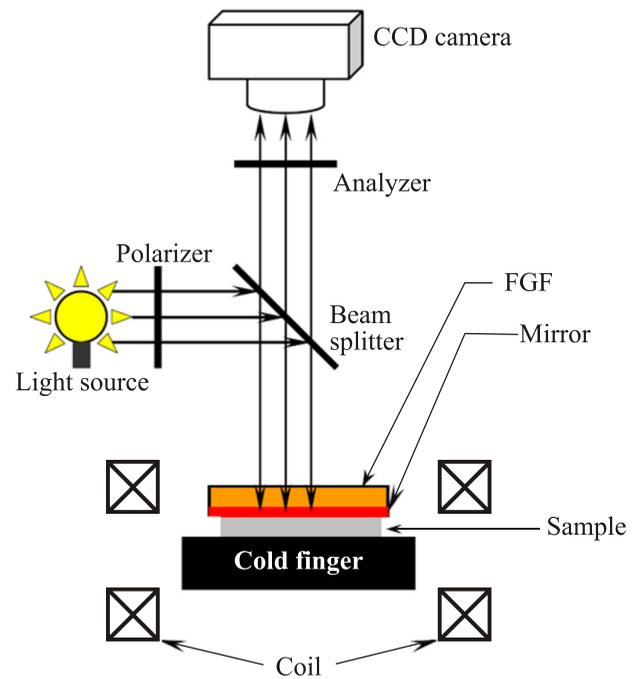

Fig. 3. Schematic of a typical MOI setup. A sample is mounted on a cold finger of a liquid He flow cryostat. Resistive coils are used as a source of an external magnetic field. The light from a mercury lamp shines through a polarizer and is guided onto an indicator film, where it experiences Faraday rotation. The light is reflected by a mirror and passes an analyzer before hitting a CCD matrix of a computer-operated camera. From Ref. 29.

## 3. Avalanche characteristics

With the use of MOI it has been discovered that in thin films avalanches have the shape of complex branching flux structures rooted at the sample edge. Such dendritic avalanches have been observed in a wide range of materials, e.g., Pb,[33] Nb,[15] Sn,[34] Nb,[17] $YBa_2Cu_3O_{7-x}$,[21] $MgB_2$,[10] $Nb_3Sn$,[19] $YNi_2B_2C$,[23] NbN,[20] and a-MoGe.[35]

From the experimental data collected on the subject (also reviewed in Ref. 29) one can identify some common features for avalanche behavior:

(i) It occurs below a certain temperature $T_{th} < T_c$.
(ii) It occurs in a limited range of applied fields: $H_1^{th} < H \leq H_2^{th}$, where $H_1^{th}$ and $H_2^{th}$ are the so-called lower and upper threshold fields, respectively.
(iii) The formation of the thermomagnetic instability is a stochastic process. Usually indentations on the sample edges serve as the most probable origins of the avalanches. Nevertheless, the exact nucleation place of the next dendrite, field interval between two consecutive events, and the final shape of the dendritic structure are essentially unpredictable.
(iv) The degree of branching of the dendritic structures, sometimes represented by their fractal dimensionality, and size vary with temperature and the applied magnetic field.
(v) Avalanches are suppressed by a metal stripe deposited along the film edge,[36,37] and deflected when meeting such strips inside the sample area.[13,38–40]

Suppression of avalanches is possible also when the metal and sample is not in thermal contact, due to the inductive braking effect.[41]

Figure 4 illustrates typical behaviors of the dendrites in a NbN film at different temperatures. At $T=4$ K the number of the dendritic avalanches per interval of the field was higher compared to $T=6$ K. The size of the dendrites shows opposite trend—it increases when the temperature approached $T_{th}$.

Criteria for onset of the thermomagnetic instability were first considered for bulks under adiabatic conditions.[42–44] The theory was later extended to include also the flow of heat,[45–48] and it was found that the instability onset can be accompanied by oscillations in temperature and electric field.[49–51] The early theory for flux jumps was reviewed in Ref. 52, see also Ref. 53. A theory for nucleation and evolution of avalanches was also developed for thick films and foils.[15]

More recent works have focused on developing theory for films placed in perpendicular magnetic field. The criteria for the instability onset were obtained from the linear stability analysis of small coordinate-dependent perturbations, focusing on edge indentations,[54,55] adiabatic condition,[56] fingering instability[57,58] and oscillatory instability.[59,60] The theory for magnetic braking as a mechanism for suppression of avalanches was also considered in Ref. 59.

When it comes to the evolution of avalanches one must rely on numerical solutions of the governing equations. Such numerical simulations have demonstrated dendrtitic avalanche behaviors with striking similarity to experimental observations[55,61,62] also revealing the ultra-fast dynamics.[63] Suppression of avalanche propagation by an adjacent metal layer was also demonstrated in simulations.[64]

## 4. Theory: Stability of metal coated thin superconductors

### 4.1. Model

Let us consider a superconducting strip of width $w$ coated with a metal layer, as depicted in Fig. 5. We assume that there is no thermal coupling between the superconductor and the normal metal, while at the same time the superconductor is thermally coupled to the substrate, which is at constant temperature $T_0$. Then the sheet current **J** consists of two contributions,[65]

$$\mathbf{J} = \mathbf{J}_s + \mathbf{J}_m, \quad (3)$$

where $\mathbf{J}_s$ and $\mathbf{J}_m$ are the sheet currents in the superconductor and metal layer, respectively. As a further approximation we assume that the electric field, **E**, is the same in the two layers, giving

$$\mathbf{J}_s = d_s \sigma_s \mathbf{E} \quad \mathbf{J}_m = d_m \sigma_m \mathbf{E}. \quad (4)$$

The thickness of the metal, $d_m$, and superconductor, $d_s$, are both much smaller than the strip width, $2w$. The conductivity of the normal metal, $\sigma_m$, is assumed to be $E$-independent, whereas the current-voltage relation in the superconducting film is assumed to be non-Ohmic with $E$-dependent conductance expressed as[66,67]

$$\sigma_s = \frac{1}{\rho_n} \begin{cases} (Ed_s/\rho_n J_c)^{1/n-1}, & J < J_c \text{ and } T < T_c, \\ 1, & \text{otherwise.} \end{cases} \quad (5)$$

Here $T$ is the local temperature, $J_c = dj_c$ is the sheet critical current of the superconductor, $\rho_n$ is the resistivity of the superconductor in the normal state, and $n$ is the creep exponent of the superconductor.

The critical current is a decreasing function of temperature, and to quantify the temperature dependence it is convenient to introduce the parameter $T^*$, defined by

$$1/T^* \equiv |\partial \ln J_c / \partial T|. \quad (6)$$

The electrodynamics is governed by the Maxwell equations in the eddy current approximation, ignoring the displacement field. The equations are

$$\nabla \times \mathbf{E} = -\dot{\mathbf{B}}, \quad \nabla \cdot \mathbf{B} = 0, \quad \nabla \times \mathbf{H} = \mathbf{J}\delta(z), \quad (7)$$

with $\mathbf{B} = \mu_0 \mathbf{H}$ and $\nabla \cdot \mathbf{J} = 0$. Due to the current conservation, it is convenient to work with the current stream function $g$ defined by Brandt[68]

$$\mathbf{J} = \nabla \times \hat{z}g. \quad (8)$$

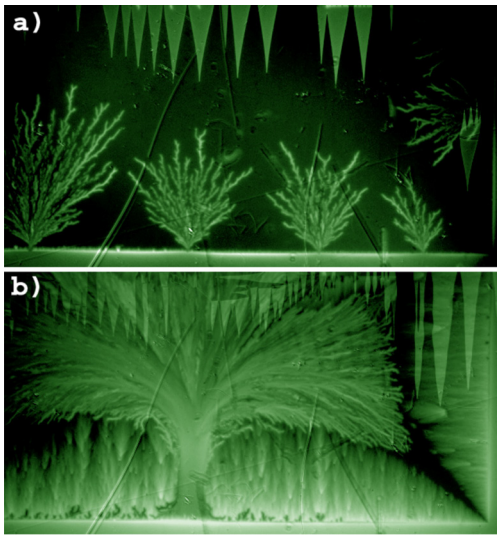

Fig. 4. Magneto-optical images of dendritic flux avalanches in a NbN film taken at (a) $T=4$ K and (b) $T=6$ K. The zigzag patterns are domain boundaries in the FGF. From Ref. 29.

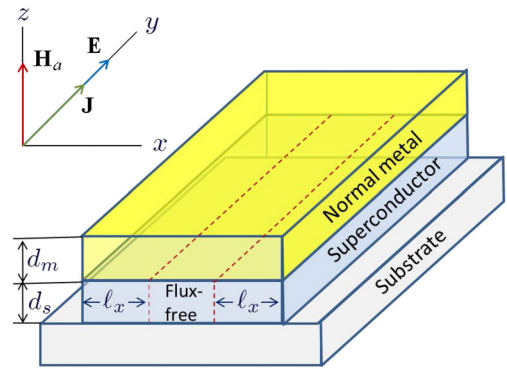

Fig. 5. Sketch of the system: a thin superconducting strip of thickness $d_s$ with a deposited metal layer of thickness $d_m$. The superconductor is in thermal contact with the substrate, kept at constant temperature $T_0$, but not with the metal. Current flows in the $y$ direction and flux has penetrated a distance $\ell_x$ from both sides due to the applied magnetic field $H_a$.

Outside the sample, $g \equiv 0$. The integral of $g$ gives the magnetic moment, $m = \int d^2r g(\mathbf{r})$. Therefore $g$ plays the role of local magnetization.

The 3D version of Ampère's law (or the Biot–Savart law) can be transformed to an integral relation in 2D.[68] In the short wavelength limit the relation has a particular simple and usefull formulation in Fourier space,

$$H_z(\mathbf{k}) = \frac{k}{2} g(\mathbf{k}), \quad (9)$$

where $\mathbf{k} = (k_x, k_y)$ are Fourier space wave-vectors.

The flow of heat in the superconductor is described by the energy balance equation describing the interplay between Joule heating, thermal conduction along the film, and heat transfer to the substrate. It reads as

$$c\dot{T} = \kappa \nabla^2 T - \frac{h}{d_s}(T - T_0) + \frac{1}{d_s} \mathbf{J}_s \cdot \mathbf{E}, \quad (10)$$

with superconductor specific heat $c$, heat conductivity $\kappa$, coefficient of heat transfer to substrate $h$. Since there is no thermal contact between the metal and the superconductor there is no need to calculate the flow of heat in the normal metal.

For further analysis it is convenient to express the equations in a dimensionless form. We let $\tilde{T} = T/T_c$, $\tilde{J} = J/J_{c0}$, $\tilde{J}_c = J_c/J_{c0}$, $\tilde{H} = H/J_{c0}$, $\tilde{x} = x/w$, $\tilde{y} = y/w$, $\tilde{t} = t\rho_n/\mu_0 d_s w$, $\tilde{E} = E/\rho_n j_{c0}$, $\tilde{\sigma}_s = \sigma_s/\rho_n$, $\tilde{\sigma}_m = \sigma_m \rho_n d_m/d_s$. Here $J_{c0} = J_c(T=0)$. Henceforth we omit the tildes for brevity.

In these units the material relations become

$$J_s = \begin{cases} J_c(E/J_c)^{1/n}, & J < J_c \text{ and } T < 1, \\ E, & \text{otherwise}, \end{cases}$$
$$J_m = \sigma_m E. \quad (11)$$

and the Maxwell equations

$$\nabla \times \mathbf{E} = -\dot{\mathbf{H}}, \quad \nabla \cdot \mathbf{H} = 0, \quad \nabla \times \mathbf{H} = \mathbf{J}\delta(z), \quad (12)$$

with $\nabla \cdot \mathbf{J} = 0$.

The heat propagation equation becomes

$$\dot{T} = \alpha \nabla^2 T - \beta(T - T_0) + \gamma J_s E, \quad (13)$$

where $\alpha$ is dimensionless heat conductivity, $\beta$ is dimensionless constant for heat transfer to the substrate, and $\gamma$ is the Joule heating parameter. The dimensionless parameters are related to the physical parameters by

$$\alpha = \frac{\mu_0 d \kappa}{\rho_n c w}, \quad \beta = \frac{\mu_0 w h}{\rho_n c}, \quad \gamma = \frac{\mu_0 w d j_{c0}^2}{T_c c}. \quad (14)$$

### 4.2. Stability analysis of bare superconductor film

Let us assume that we start from uniform background distributions of the electric field $\mathbf{E} \equiv E\hat{\mathbf{y}}$ and temperature $T$, as depicted in Fig. 5. The left edge of the sample is at $x = 0$, the right is at $x = 2$. Due to the applied magnetic field or current, the magnetic flux front, and thus also the fronts of $E$ and $T$ have reached a distance $l_x$ from both edges. The perturbed values are specified as $\mathbf{E} + \delta\mathbf{E}$, $T + \delta T$, etc. To meet the boundary conditions we assume that in the Fourier space the perturbations are of the form

$$\delta T \propto e^{\lambda t} \cos(k_x x) \cos(k_y y),$$
$$\delta J_x, \delta E_x \propto e^{\lambda t} \sin(k_x x) \sin(k_y y),$$
$$\delta J_y, \delta E_y \propto e^{\lambda t} \cos(k_x x) \cos(k_y y),$$
$$\delta H_z \propto e^{\lambda t} \sin(k_x x) \cos(k_y y), \quad (15)$$

where $\lambda$ is the instability increment and $k_x$ and $k_y$ are the in-plane wave-vectors. The flux penetration depth sets the lower limit for allowed wave-vectors in $x$ direction and we will thus identify $l_x = \pi/2k$ and let the corresponding $l_y = \pi/2k_y$ be determined by the analysis. We will now linearize the equations in the perturbations and find the eigenvalue equation for the instability increment, $\lambda$.

The onset of instability typically happens at low electric fields, when all current flows in the superconductor and nothing in the metal. We thus let

$$J_m = 0. \quad (16)$$

We further assume that $n \gg 1$, $J = J_c$, and $T = T_0$.

The eigenvalue equation for the instability increment $\lambda$ was derived in Ref. 59. It can be cast in the form

$$A\lambda^2 + B\lambda + C = 0, \quad (17)$$

where

$$A = \frac{k}{2}\frac{J_c}{nE},$$
$$B = k_x^2 + \frac{k_y^2}{n} + \frac{k}{2}\frac{\alpha k^2 + \beta}{nE} J_c - \frac{k}{2}\frac{\gamma J_c}{T^*},$$
$$C = (\alpha k^2 + \beta)\left(k_x^2 + \frac{k_y^2}{n}\right) + (k_x^2 - k_y^2)E\frac{\gamma J_c}{T^*}. \quad (18)$$

In order to find the instability threshold conditions we must solve for Re $\lambda = 0$.

Let us first consider the case when $\lambda$ is real. The instability onset condition $\lambda = 0$ then implies that

$$C = 0. \quad (19)$$

From Eq. (18) we see that $C = 0$ corresponds to the case when $l_x \gg l_y$ and this case is therefore often called a fingering instability.[48,57] The most unstable mode is determined by $\partial \lambda/\partial k_y = 0$, giving $\partial C/\partial k_y = 0$. Eliminating $y$ and solving for $E$ gives the threshold electric field for the fingering instability

$$E_{\text{th}}^{\text{Fingering}} = \frac{T^*}{\gamma J_c}\left(\sqrt{\alpha}k_x + \sqrt{\frac{\beta}{n}}\right)^2. \quad (20)$$

This expression was also considered in Refs. 57, 58, 69, and 70.

Let us next consider the case when $C > 0$. In this case $\lambda$ is complex and the instability threshold is determined by the condition Re $\lambda = 0$, which yields

$$B = 0. \quad (21)$$

This corresponds to a solution with temporal oscillations with frequency

$$\omega = \sqrt{C/A}. \quad (22)$$

Also in this case, the most unstable mode is found by the condition $\partial \operatorname{Re} \lambda / \partial k_y = 0$, which gives $\partial B / \partial k_y = 0$.

Again we refer to Ref. 59 for the calculations. They lead to the following expression for the threshold electric field,

$$E_{\text{th}}^{\text{Oscillatory}} = \frac{\beta T^*}{\gamma J_c n} (u_+ + u_-)^{-3}, \quad (23)$$

where

$$u_\pm = \left[ \frac{1}{2} \pm \sqrt{\frac{1}{4} + \frac{\alpha}{\beta} \left( \frac{T^* k_x^2}{\gamma J_c^2} \right)} \right]^{1/3}.$$

Series expansion in $k_x$ gives

$$E_{\text{th}}^{\text{Oscillatory}} = \frac{T^* \beta}{\gamma J_c n} \left[ 1 + 3 \left( \frac{\alpha}{\beta} \right)^{1/3} \left( \frac{T^* k_x^2}{J_c^2 \gamma} \right)^{2/3} \right]. \quad (24)$$

The peculiar $k_x^{4/3}$ dependence is due to the $k/2$ Fourier kernel.

Equations (23) and (24) are rather complicated, therefore it is practical to approximate them. A relatively simple approximation can be obtained in the limit of $l_y = \infty$, which implies that the instability is uniform. From $C = 0$ in Eq. (18) one gets

$$E_{\text{th}}^{\text{Uniform}} = \frac{J_c}{n} \frac{\alpha k_x^2 + \beta}{\gamma J_c^2 / T^* - 2 k_x}. \quad (25)$$

The physical interpretation of Eq. (25) is straightforward: increasing heat removal through $\alpha$ and $\beta$ leads to increase of the threshold, while increasing Joule heating through $\gamma$ and non-linearity through $n$ leads to its decrease. In the extreme Bean model limit, $n \to \infty$, the threshold is independent of $E$, $\alpha$ and $\beta$ and the threshold condition is purely adiabatic, $k_x = \gamma J_c^2 / 2T^*$. This case was considered also in Ref. 56.

Let us now compare the three expressions Eqs. (20), (23), and (25) for the threshold electric field. Figure 6 shows temperature dependences of the critical electric fields corresponding to the fingering, fingering oscillatory and uniform oscillatory types of the instability, Eqs. (20), (23), and (25), respectively. For the plots we assumed constant $\alpha$ and $\beta$, and the temperature dependences $J_c = 1 - T$, $n = n_1/T$ and $\gamma = \gamma_0 T^{-3}$, where $\gamma_0$ is constant. The figure shows that threshold fields for the oscillatory cases are significantly lower at most temperatures. Therefore, the oscillatory modes will most likely initiate the instability. The plot also shows that Eq. (25) is good approximation for Eq. (23) for low $T$.

### 4.3. Reentrant stability due to magnetic braking effect

Let us now consider the case when electric field is high, i.e., an avalanche is already progressing. When the superconductor is covered by normal metal the electromagnetic braking effect may open the possibility of reentrant stability at high electric field. A practical consequences of this reentrant stability is that an avalanche may stop at an early stage before much damage has been done.

For the analysis, it is convenient to introduce the nonlinearity exponent of the composite system as

$$n_{\text{tot}}(T, E) \equiv \frac{\partial \ln E}{\partial \ln J} = n \frac{1 + J_m/J_s}{1 + n J_m/J_s}. \quad (26)$$

The magnetic braking is strong when $n_{\text{tot}} \sim 1$.

The linear stability analysis of the composite system was carried out in Ref. 59. Also in this case the eigenvalue equation of $\lambda$ was quadratic, but the factors were more complicated than for the uncoated sample. The eigenvalue equation is

$$A\lambda^2 + B\lambda + C = 0, \quad (27)$$

with

$$A = \frac{k}{2} \frac{J}{n_{\text{tot}} E},$$

$$B = k_x^2 + \frac{k_y^2}{n_{\text{tot}}} + \frac{k}{2} \frac{\alpha k^2 + \beta}{n_{\text{tot}} E} J - \frac{k}{2} \left( \frac{J_s}{J_c} - \frac{1}{n_{\text{tot}}} \right) \frac{J_s}{J_c} J \frac{\gamma J_c}{T^*},$$

$$C = (\alpha k^2 + \beta) \left( k_x^2 + \frac{k_y^2}{n_{\text{tot}}} \right) + \left[ k_x^2 - k_y^2 \left( \frac{J_s}{J} - \frac{1}{n_{\text{tot}}} \right) \right] \frac{J_s}{J_c} E \frac{J_c \gamma}{T^*}. \quad (28)$$

The form-factor of the avalanche at high electric field is in general difficult to predict as it is a consequence of the nonlinear and nonlocal evolution of the instability. Consequently it is difficult to constraint $k_x$ and $k_y$. However, assuming that the avalanche is at an early stage of development, the form-factor should be pretty much the same as for the onset of instability, and then the most unstable mode typically have $k_x > k_y$ and this implies that also in this case that the oscillatory modes are most relevant, and we should consider $B = 0$ as the condition for reentrant stability.

In the limit when $n J_m \gg J_s$ we have

$$n_{\text{tot}} \approx 1 + J_s/J_m, \quad (29)$$

where $J_m = \sigma_m E$ and $J_s \approx J_c$, when $n \gg 1$. Using this in the condition $B = 0$ leads to the condition for reentrant stability by magnetic braking as

$$B = k_x^2 + \frac{k_x}{2}(\alpha k_x^2 + \beta)\sigma_m - \frac{k_x}{2}(J_c - \sigma_m E)\frac{\gamma J_c}{T^*} = 0. \quad (30)$$

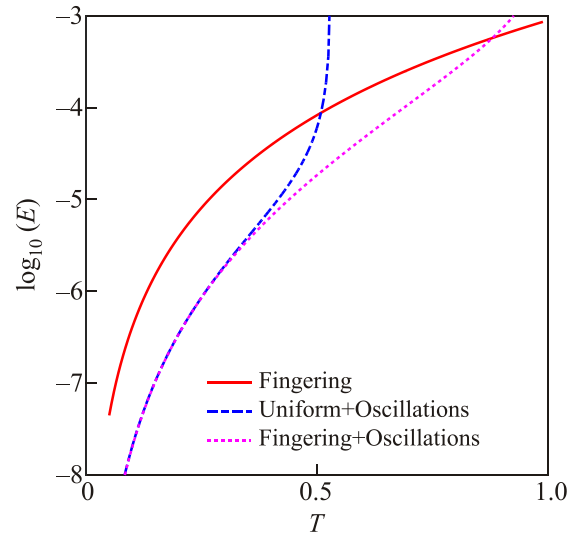

Fig. 6. The threshold for onset of instability in the $T$–$E$ plane, for the fingering, fingering + oscillatory, and uniform + oscillatory conditions. In a uniform sample, the lowest of these curves determines the onset of instability. The parameters are $\alpha = 10^{-5}$, $\beta = 0.1$, $\gamma_0 = 10$, $l_x = 0.1$, $n_1 = 50$.

Solving for $E$ gives

$$E = \frac{1}{\sigma_m}\left(J_c - \frac{2k_x}{F}\right) - \frac{\alpha k_x^2 + \beta}{F}. \quad (31)$$

The reentrant stability thus appearing at high electric fields, of the order of $E \sim J_c/\sigma_m$.

Shown in Fig. 7 are the stability diagrams in the $T$–$E$ plane for different conductivity of the coating metal. The curves have been calculated by numerical solution of Eq. (27). They demonstrate that metal coating increases stability of the flux distribution. In particular we see that stability reappears at high electric fields, typically of order $E \sim J_c/\sigma_m$. From the figure we also see that it is possible draw a connected path between the stable configurations at high and low electric fields. This opens the possibility that avalanches in coated regions can stop and reenter the low-$E$ state.

## 5. Simulation: Evolution of avalanches in metal coated sample

### 5.1. Procedure

Considering a type-II superconducting thin film in transverse applied field, we will now describe our scheme for numerical simulations of the flux dynamics. The inputs for the simulations are the nonlinear $E$–$J$ relations characterizing the material properties of the films and the ramping of the external magnetic field, $\dot{H}_a$. In order to carry out such simulations one must overcome the problem of imposing the boundary conditions. This is challenging due to the inherent self-induction of the system. One way to handle the overcome the self-induction problem is to include the sample boundary directly in the discretization of the sample. Brandt has invented a series of such discretization schemes for, e.g., squares and rectangles,[68] disks and rings,[71] and arbitrary connected geometry.[72] An alternative, approximate and much more numerically efficient approach is to discretize without taking into account the sample boundaries and instead impose the boundary conditions indirectly through a real-space Fourier-space hybrid method. This approach has been used for a series of geometries.[61,62,73]

We will now consider the case of a superconducting film partly covered by metal and simulate the evolution of a dendritic flux avalanche to find the effect of magnetic braking on the evolution of the avalanche. The description uses the same dimensionless units as used in the linear stability analysis. We adopt the model of Eq. (4) were the superconductor–metal composite system is considered as two conductors connected in parallel,

$$\mathbf{E} = (\sigma_s + \sigma_m)^{-1}\mathbf{J}, \quad (32)$$

where $\sigma_m$ is constant conductivity of the metal layer. The nonlinear superconductor conductivity is given in Eq. (5) as $\sigma(E)$ but for simulations we need $\sigma(J)$ and the inversion cannot be expressed in a closed form. Instead we use

$$\frac{1}{\sigma_s} = \begin{cases} (J/J_c)^{n+1}, & T < T_c \text{ and } J < J_c, \\ 1, & \text{otherwise,} \end{cases} \quad (33)$$

where $J_c$ is the critical sheet current and $n \gg 1$ is the creep exponent. In Eq. (33) we have used the total sheet current rather than the part flowing in the superconductor. This is a good approximation when $\sigma_m E \ll J$, like during the regular flux penetration, and in the very initial stage of an avalanche. During the propagation stage of an avalanche the $E$-field is large, and our simplification leads to underestimation of the magnetic braking effect.

The numerical simulations are most conveniently formulated using the local magnetization, $g$, defined in Eq. (8). For quasi-static situation $H_z$ is the superposition of the applied field and film self-field. Using Eq. (9) we write

$$H_z = H_a + \hat{Q}g, \quad (34)$$

with the operator $\hat{Q}$ given by

$$\hat{Q}g(\mathbf{r}) = \mathcal{F}^{-1}\left[\frac{k}{2}\mathcal{F}[g(\mathbf{r})]\right], \quad (35)$$

where $\mathcal{F}$ is the 2D spatial Fourier transform, $k = |\mathbf{k}|$, and $\mathbf{k}$ is the wave-vector. The inverse relation is

$$\hat{Q}^{-1}\varphi(\mathbf{r}) = \mathcal{F}^{-1}\left[\frac{2}{k}\mathcal{F}[\varphi(\mathbf{r})]\right], \quad (36)$$

where $\varphi$ is an auxiliary function.

By taking the time derivative of Eq. (34) and inverting it, we get

$$\dot{g} = \hat{Q}^{-1}[\dot{H}_z - \dot{H}_a]. \quad (37)$$

This equation is solved by discrete integration forward in time.

Regarding the discretization of space, the key point is that both $\hat{Q}$ and $\hat{Q}^{-1}$ are direct products in Fourier space which means that the operators can be calculated effectively using Fast Fourier Transforms (FFT). However, the derivation leading to the simple form for $\hat{Q}$ and $\hat{Q}^{-1}$ has neglected the sample boundary, which means that also the vacuum surrounding the sample must be explicitly included in the calculations. The total area of calculations is thus a rectangle of dimensionless $L_x \times L_y$ including both sample and vacuum. The solution will be periodic on this larger rectangular area.

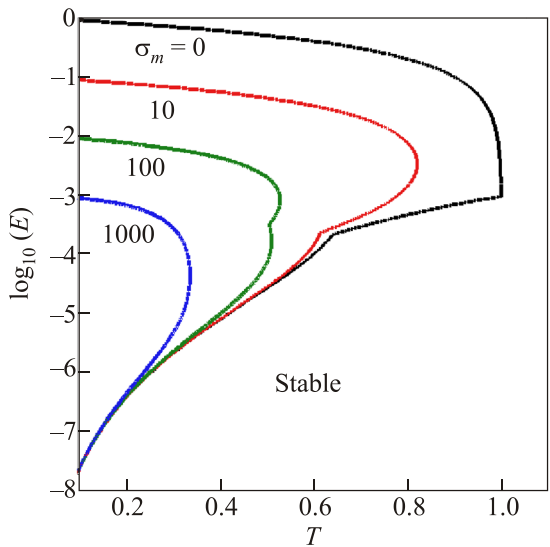

Fig. 7. The lines show the boundary of the instability region when changing the normal metal conductivity $\sigma_m = 0$, 10, 100, and 1000. Increasing metal layer conductivity improves stability at high $E$ and $T$. Parameters are $\alpha = 10^{-5}$, $\beta = 0.1$, $\gamma = 10$, $l_x = 0.1$, $n_1 = 20$.

Thus, in order to integrate Eq. (37) forward in time, $\dot{H}_z$ must be known everywhere in the embedding $L_x \times L_y$ rectangle at time $t$. Our strategy is to find $\dot{H}_z$ inside the sample from the material law, while in the vacuum $\dot{H}_z$ is found implicitly from the condition $\dot{g} = 0$, as described below.

Starting with the superconductor itself, it obeys the material law, Eq. (32), which, when combined with the Faraday law from Eq. (12), gives

$$\dot{H}_z = \nabla \left( \frac{\nabla g}{\sigma_s + \sigma_m} \right). \quad (38)$$

From $g(\mathbf{r},t)$ the gradient is readily calculated, and since the result allows finding $\mathbf{J}(\mathbf{r},t)$ from Eq. (8) also $\sigma_s(\mathbf{r},t)$ is determined from Eq. (33).

The task then is to find $\dot{H}_z$ outside the sample boundaries so that $\dot{g} = 0$ outside the superconductor. This cannot be calculated efficiently using direct methods due to the nonlocal $\dot{H}_z - \dot{g}$ relation and the non-symmetric sample shape. Instead we use an iterative procedure.

For all iteration steps, $i = 1\ldots s$, $\dot{H}_z^{(i)}$ is fixed inside the superconductor by Eq. (38). At is $i = 1$, an initial guess made for $\dot{H}_z^{(i)}$ outside the sample, and $\dot{g}^{(i)}$ is calculated from Eq. (37). In general, this $\dot{g}^{(i)}$ does not vanish outside the superconductor, but an improvement can be obtained by

$$\dot{H}_z^{(i+1)} = \dot{H}_z^{(i)} - \hat{Q}\hat{O}\dot{g}^{(i)} + C^{(i)}. \quad (39)$$

The projection operator $\hat{O}$ is unity outside the superconductor and zero inside. To improve the numerical stability one should shift $\hat{O}\dot{g}^{(i)}$ to satisfy $\int d^2r \hat{O}\dot{g}^{(i)} = 0$. The constant $C^{(i)}$ is determined by requiring flux conservation,

$$\int d^2r \left[ \dot{H}_z^{(i+1)} - \dot{H}_a \right] = 0. \quad (40)$$

Thus, at each iteration $i$, $\dot{H}_z^{(i+1)}$ is calculated for the outside area. The procedure is repeated until after $i = s$ iterations $\dot{g}^{(s)}$ becomes sufficiently uniform outside the sample. Then, $\dot{g}^{(s)}$ is inserted in Eq. (37), which brings us to the next time step, where the whole iterative procedure starts anew.

The state is numerically described by $g$ and $T$. The time evolution are obtained by simultaneous time integration of Eqs. (37) and (13).

### 5.2. Simulation result

Let us now consider the time evolution of partly metal coated sample. The metal layer is considered to be thermally isolated from the superconductor, and the only effect of the metal layer is the magnetic braking at high electric fields. The theory of Sec. 4.3 predicts that the superconductor can enter a regime of stability at high electric fields and this may lead to a suppression of the avalanches in the metal coated parts.

The sample is a superconducting square where the right half is covered by a metal of high conductivity, $\sigma_m = 1000$ The parameters of the simulation are $n_1 = 20$, $\alpha = 10^{-5}$, $\beta = 0.07$, $\gamma_0 = 10$ and $\dot{H}_a = 10^{-8}$. The simulation procedure was carried out in two steps. First, the flux penetration was simulated at constant temperature. Second, the state was rescaled to account for finite temperature,[62] temperature was allowed to vary, and a avalanche was nucleated by a heat pulse slightly off-center, in the non-metal-covered part. We then follow the evolution of the avalanche.

Figure 8 shows the distributions of $H_z$, $T$, and $J$ at times $t = 0.25$, 12.25 and 24.75 after nucleation of the avalanche. The blue, stippled line in the figure marks the edge of the metal cover.

At $t = 0.25$ the avalanche is just a narrow finger barely extending the critical state region. It has already at this stage reached a temperature above $T_c = 1$. Note that the thickness of the finger is determined by the propagation of the hot spot and is not related to the size of the thermo-magnetic instability at nucleation stage.[59] At $t = 12.25$ the avalanche has the characteristic branching shape typically observed by magneto-optical imaging at times after the avalanches has stopped propagating.[10] Yet, this avalanche is still propagating and the branches are heated above $T_c = 1$. Flux has accumulated at the boundary of the metal cover and we see that protection is almost complete as the avalanche does not propagate into the metal covered part. At $t = 24.75$ the avalanche is close to its final extent. The temperature now is 0.5 and decaying. There is a minor inclusion of the avalanche into the metal covered part, but the protection offered by the metal is good. The level of the shielding currents at the boundary is high—comparable with the critical state region. Yet, the maximum magnitude of the current is lower that $t = 12.25$, since the strong eddy currents in the metal layer decays on the time scale comparable with the time scale of the avalanche.

### 6. Ultra-fast propagation of avalanches

The avalanche events occur unpredictably and develop too fast to be followed dynamically by any experimental method available today. With conventional magnetometry one observes only a step in the magnetic moment due to the abrupt redistribution of flux and induced currents.[41,74] More information is obtained from magneto-optical imaging (MOI), where the spatial distribution of magnetic flux before and after the breakdown is visualized using a Faraday-active sensor mounted on the sample. However, results providing insight into how the breakdown evolves in time are extremely scarce. Only by using a femtosecond pulsed laser to actively trigger an event it was possible to synchronize the image recording and to capture the flux distribution at an intermediate stage.[21,22,75,76] From those experiments it was found that the flux front can advance at an astonishing speed exceeding 100 km/s. This ultra-fast dynamics causes a lot of questions, which we have addressed by performing numerical simulations of the thermo-electromagnetic behavior of an uncoated superconducting thin circular disk,[63] see Fig. 9, using material parameters corresponding to superconducting $MgB_2$. A magnetic field $H_a$ is applied transverse to the sample plane, and as it gradually increases from zero it drives the penetration of magnetic flux into the disk. In the early stage of the field ramp, the flux enters evenly around the edge, and advances to increasing depth without any sign of intermittent behavior. In the penetrated region a critical state is formed and characterized by a sheet current $J$ and flux density $B_z$ in full agreement with the Bean model for a thin circular disk.[8,9,71]

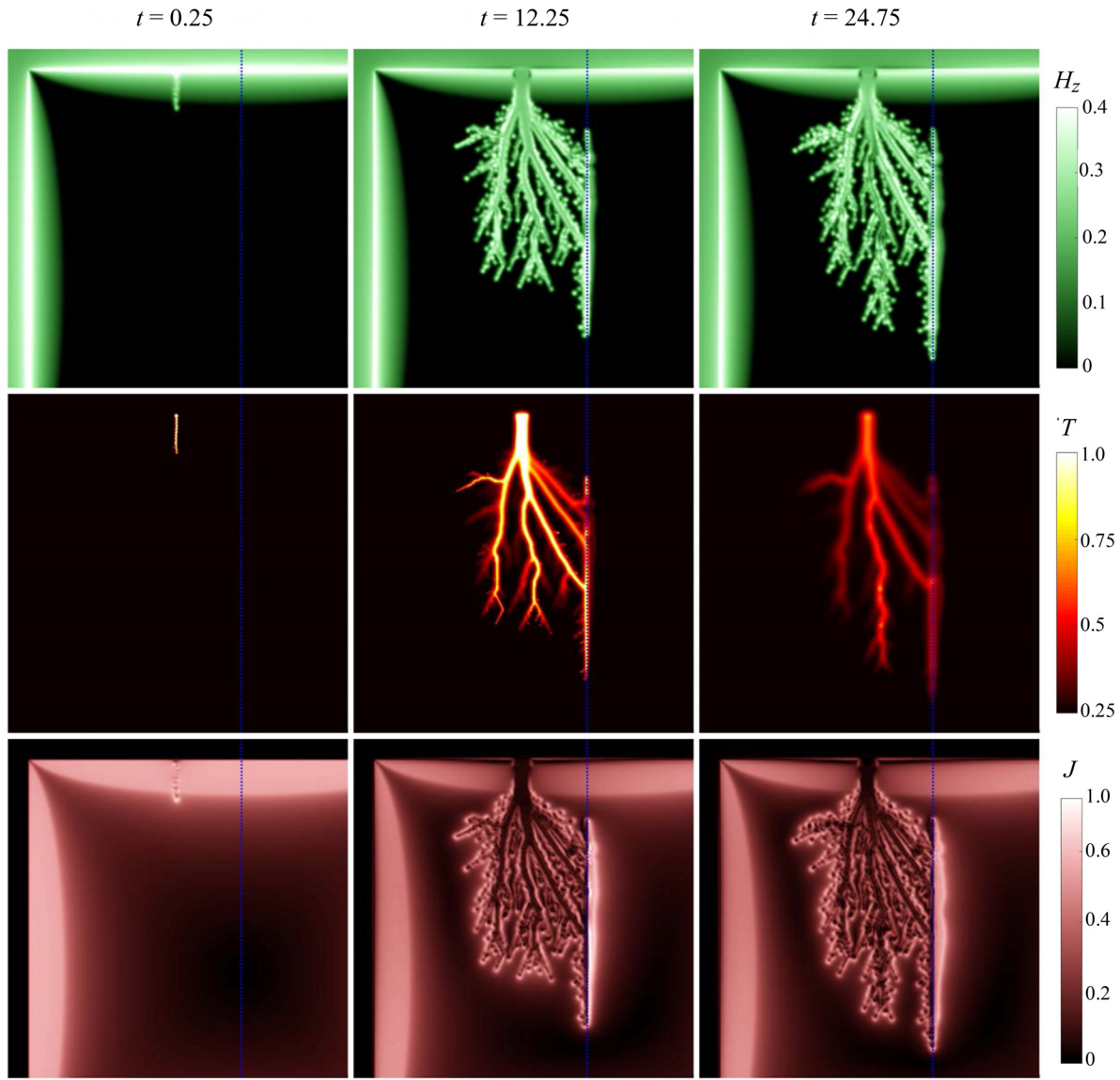

Fig. 8. Simulated evolution of an avalanche in a sample where the region to the right of the dotted line is covered with metal with $\sigma_m = 1000$. Distributions of the magnetic field $H_z$, temperature $T$ and sheet current magnitude $J$, at times $t = 0.25$, 12.25 and 24.75 after nucleation of the avalanche.

In our calculations we focused on the temporal evolution of the flux pattern, which is beyond experimental accessibility. When the applied field reaches $\mu_0 H_{th} = 5.3$ mT the first abrupt event is nucleated, and magnetic flux enters from the edge. A complex branching structure is created as the flux invades deep into the flux-free region, see Fig. 10(a). As $H_a$ continues increasing, only the gradual flux the dendritic structure remains frozen. Then, at the field of 6.2 mT, another similar event takes place in a different part of the sample, and soon thereafter yet another one strikes.

In this way the superconductor experiences a sequence of dramatic events at unpredictable intervals and locations, and where each breakdown follows an intriguing path in a macroscopically uniform medium. Since this phenomenon is of electrodynamic nature, it is interesting to recognize the many aspects that are similar to atmospheric lightening. Figure 10(b) shows MOI picture of the flux distribution in a superconducting MgB$_2$ film at $T = 5$ K where the magnetic field had been increased from zero to $\mu_0 H_a = 3.8$ mT. The experimental image reveals that the flux avalanches have a morphology quite similar to the numerical results, and also that the events have a clear tendency to avoid spatial overlap, as in the simulations.

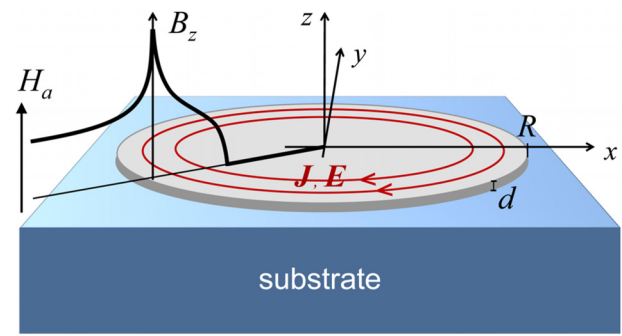

Fig. 9. Sample configuration. A thin superconducting disk on a substrate exposed to a gradually increasing perpendicular magnetic field, $H_a$. The flux density, $B_z$, is advancing from the edge along with a distribution of induced shielding cur-rent, $J$, and electrical field, $E$. From Ref. 63.

To analyze time evolution of magnetic flux distribution we focus on the detailed dynamics of one breakdown, and we choose to zoom in on the event taking place at $\mu_0 H_a = 5.3$ mT. Shown in Fig. 11 rows (a)–(d) are five instantaneous distributions of the magnetic flux density $B_z$, the stream line pattern of the flow of sheet current $J$, the temperature $T$, and the electric field $E$, respectively. The snapshots show the states at $t = 1, 5, 22, 52$ and $86$ ns after nucleation of the instability. The final flux distribution looks quite similar to those reported from many MOI experiments.[10,11,13,16,17,19–21,23,37,75,77–81] The reported high velocities of the flux propagation are also confirmed.

Our simulations have revealed several important time scales characterizing the nucleation and subsequent evolution of the thermo-electromagnetic breakdown in superconducting films. First, we find that the rise time of such events, described by how fast the electric field rises to its maximum, is extremely short: of the order of 1 ns. The total duration of an event is 75–80 ns, while the nucleation of a new branch takes less than 5 ns (Fig. 11).

The shortest time scale, $\tau_a$, describes time to increase the temperature from $T_0$ to $T_c$. This characteristic time is estimated by considering Ohmic Joule heating, and solving the equation $c(T)\dot{T} = \rho_0 j_c^2(T_0)$ where $c(T) = c(T_c)(T/T_c)^3$ is the specific heat. Integrating this equation gives

$$\tau_a = c(T_c)T_c/4\rho_0 j_c^2(T_0), \quad (41)$$

where a small term $\sim (T_0/T_c)^4$ is ignored. Using the material parameters given in Ref. 63, the numerical value becomes $\tau_a = 0.5$ ns, which indeed is very close to the rise time of the simulated events.

The electromagnetic time scale, $\tau_{em}$, describes the lifetime of normal currents. For a thin disk, Brandt has found that the longest surviving mode has a decay time given by[82]

$$\tau_{em} = 0.18\mu_0 R d/\rho_0. \quad (42)$$

With the present parameters, this gives $\tau_{em} = 1.8$ ns. It worth noting that in the bulk case such a time constant cannot be defined since the flux motion is then described by a diffusion equation. In films, on the other hand, the flux penetration is accelerated by the presence of a free surface. The decay time is related to the propagation velocity of the peak in the current density, which is $v_{em} = 0.77\rho_0/\mu_0 d = 0.14 R/\tau_{em} = 140$ km/s.[82] This value provides the upper bound for the propagation velocity of the dendrite. Indeed, the initial dendrite tip velocity $\sim 90$ km/s of is not far from $v_{em}$.

Note that $v_{em}$ is proportional to the normal resistivity $\rho_0$. In the next section we will demonstrate that this property can be used for tuning the velocity by coating the superconductor by a normal metal.

Heat removal to the substrate leads to an exponential decay of the temperature with a time constant

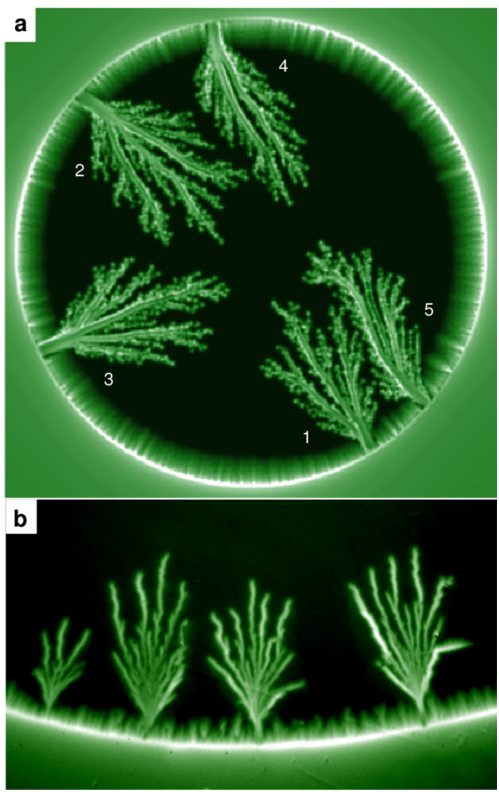

Fig. 10. Flux density after a few breakdown events. (a) Simulated distribution of $B_z$ in a superconducting disk after five flux avalanches occurred in the sequence indicated by the numbers as the applied field was ramped up from zero to $\mu_0 H_a = 8.5$ mT. (b) Magneto-optical image of the flux density in a superconducting MgB$_2$ film cooled to 6 K and then exposed to an applied field of 3.8 mT. From Ref. 63.

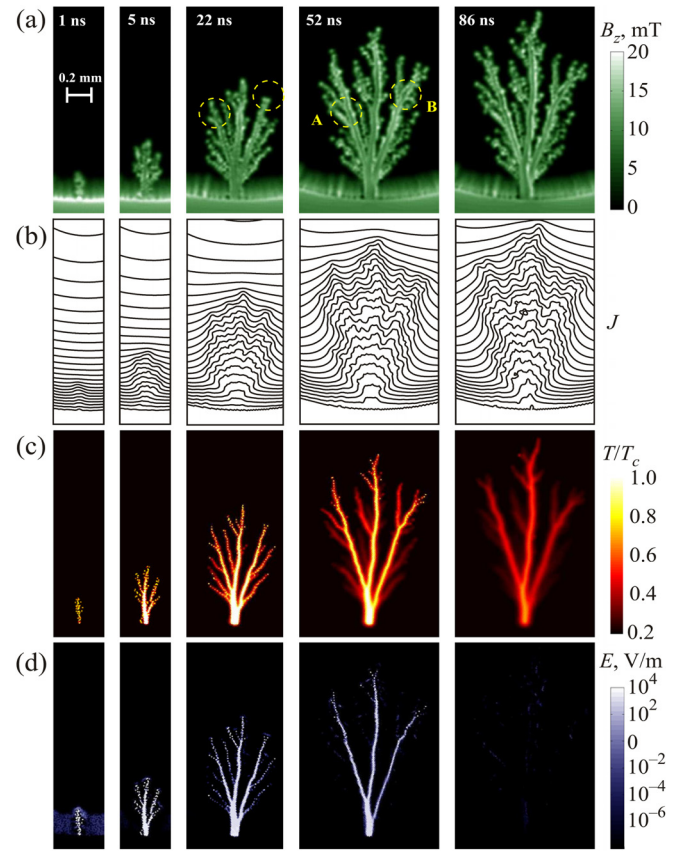

Fig. 11. Evolution of a breakdown event. (a) Distributions of the magnetic flux density $B_z$, (b) the induced sheet current $J$, (c) the temperature $T$, and (d) the electrical field $E$, at times $t = 1, 5, 22, 52$ and $86$ ns after nucleation of the thermo-electromagnetic instability.

$$\tau_h = cd/h = 52 \, \text{ns}, \quad (43)$$

where $h$ is the coefficient of heat transfer to the substrate. We find that indeed $\tau_h \gg \tau_a, \tau_{em}$. It is fully consistent with the fact that the events actually do take place, rather than being prevented by an efficient heat sink provided by the substrate. The value of $\tau_h$ is comparable to the total duration of the event, suggesting that the heat removal to the substrate largely determines the avalanche life-time, and thereby also decides the size of the full-grown flux dendrite.

Finally, the lateral heat transport is an ordinary diffusion process with diffusion time

$$\tau_\kappa = l^2 c/4\kappa, \quad (44)$$

where $l$ is the diffusion length and $\kappa$ is the thermal conductivity. The diffusion length characteristic for the dendrite tips can be obtained from the $T$-maps of Fig. 11. The very sharp tips of the growing branches have a typical width $l = 10 \, \mu$m, which gives $\tau_\kappa = 3.7$ ns. This is close to the 5 ns time when the first branching of the structure was detected, indicating that the heat diffusion should contribute to the branching process. Considering the other extreme, and letting $\tau_\kappa$ be the total duration of an event, 75 ns, we obtain the largest relevant diffusion length, $l = 125 \, \mu$m. This is much smaller than the length of the long branches in the dendritic structure, but interestingly it is approximately half the width of the dendrite trunk at the final stage. This indicates that the trunk is gradually widened by heat diffusion during the event.

Note that the time scale of the background flux penetration is on the order of milliseconds, i.e., it is much longer than the characteristic time scales estimated above. Therefore, our results on the evolution of the instability are essentially independent of the ramp rate of the applied magnetic field. This robustness is consistent with numerous MOI experiments performed by some of the present authors.

## 7. Ray optics behavior of avalanche propagation

As it was mentioned in the previous section, the propagation of the dendrite trunk is very similar to an electromagnetic wave in a normal layer, its velocity, $v_{em}$, being proportional to the metal resistivity $\rho_0$.[82,83] Therefore, one can expect that the trunks should refract at the boundaries between the regions with different effective resistivity. Indeed, previous work by Albrecht et al.[13,84] showed that the propagation of flux dendrites crossing borders between regions of different material properties depends on the incidence angle of the avalanche.

A natural way to prepare such a system is to coat the superconducting film by a normal metal with relatively high conductivity exceeding that of the superconductor material in the normal state. This idea was realized in Ref. 39 using NbN film patterned with Cu strips. Films of NbN were grown on MgO(001) single crystal substrate to a thickness of 170 nm using pulsed laser deposition. By electron beam lithography and reactive ion etching with $CF_4 + O_2$, one film was shaped into a 3.0 × 1.5 mm rectangle. Then, a 900 nm thick Cu layer was deposited on the film and patterned as shown in Fig. 12. Here, the two long horizontal strips of metal define areas where flux avalanches starting from the lower film edge will experience magnetic braking.

The metal coating along the upper edge has the purpose of preventing avalanches to start from that sample side.

In addition to MOI observations contact pads were placed at the lower corners of the sample, where the left pad contacts the two long metal strips. These contact pads were used to pick up the voltage pulses generated by flux avalanche propagating in a metal-coated part of a superconductor film.[38] With this geometry, if two subsequent pulses are detected they provide information about the speed of the avalanche front. Moreover, the fine structure of each pulse tells about the number of flux branches passing the electrodes and the points in time they enter and exit.

Shown in the upper panel of Fig. 13 is a magneto-optical image of the flux distribution after a typical avalanche occurred in the NbN film at 3.7 K in descending applied magnetic field. Prior to the field descent, the film was filled with flux by applying a perpendicular field of 17 mT, which removed essentially all the flux trapped from previous experiments, and created an overall flux distribution corresponding to a critical state. Then, during the subsequent field descent, when the field reached 14 mT, a large-scale avalanche started from a location near the center of the lower sample edge. The dark dendritic structure shows the paths followed by antiflux as it abruptly invaded the sample.

Note that as long as the ray propagation takes place in the same medium, i.e., either the bare superconductor or the metal-coated area, the rays are often quite straight. Moreover, when the rays traverse an interface between the two media, their propagation direction is changed displaying a clear refraction effect.

A magnified view of the flux distribution inside the rectangular area marked in Fig. 13 (upper) is shown in the lower panel. In the metal strip area the rays, indicated by dashed yellow lines, traverse the strip at various angles denoted $\theta_i$, see the insert for definitions. As the rays cross the interface they continue into the bare superconductor at a different angle $\theta_r$. This refraction angle is consistently larger than the incident angle, $\theta_i$, and it is interesting to compare the two angles quantitatively in relation to Snell's law,

$$\sin \theta_r / \sin \theta_i = n.$$

Here $n$ is the relative index of refraction of the metal-coated and bare areas of the superconductor. From the examples of refraction indicated by the dashed lines in Fig. 13 (lower) one finds $n = 1.37, 1.37, 1.44$ and $1.34$, which are remarkably similar values. Note that the metal strip nearest the edge is essentially invisible since it does not lead to refraction. This is fully consistent with Snell's law since the avalanche here enters the strip at normal incidence.

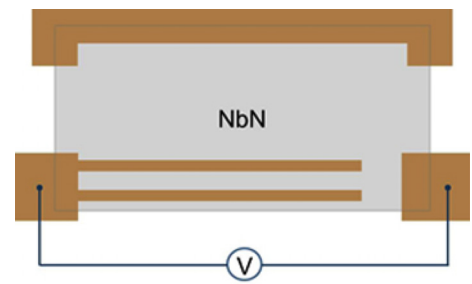

Fig. 12. Schematics of the rectangular NbN super-conducting film covered by a Cu-layer patterned as seen in the figure. Shown is also the voltage pulse measurement circuit, which allows time-resolved observation of the avalanches starting from the lower film edge. From Ref. 39.

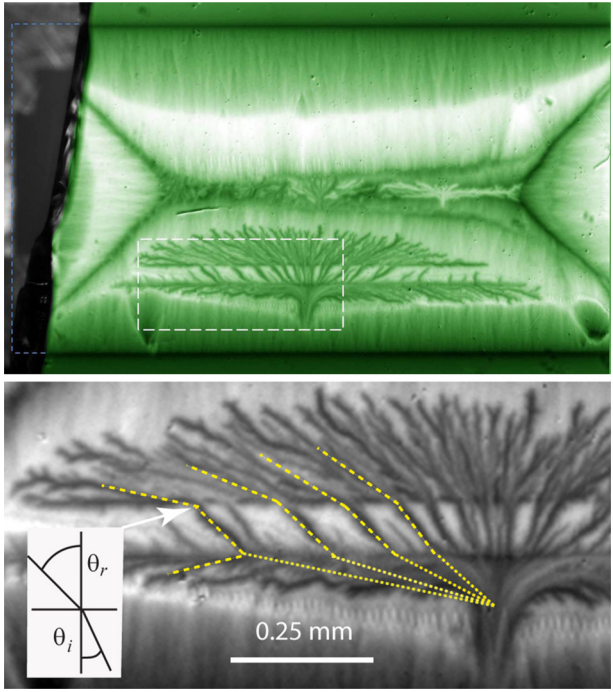

Fig. 13. Magneto-optical image of a flux avalanche occurring at 3.7 K in the metal coated NbN film. The image covers the lower central part of the film, and was recorded in the remnant state after the field was first raised to 17 mT. The horizontal bright strip permeated by dark line segments is the metal coated strip located nearest to the sample center. The strip near the edge is invisible, as the avalanche crossed this region through a single channel perpendicular to the edge. From Ref. 39.

These observations give strong indications that the avalanche dynamics is governed by oscillatory electromagnetic modes, and that these modes have different propagation velocities in the bare superconductor and metal-coated film. Denoting these two velocities $v_s$ and $v_c$, respectively, the suggested physical picture then demands that their ratio is equal to the index of refraction, $v_s/v_c = n$. This relation was tested by analyzing additional experimental data from voltage pulses between the contact pads.

The surprising observation that branches of a flux avalanche propagating across boundaries between two superconducting media show quantitative agreement with Snell's refraction law. This leads us to conclude that the branches propagate as electromagnetic modes with well-defined speed. Such modes propagating in a film of resistivity $\rho$ were considered in Refs. 82 and 83 where it was found that their speed can be written as

$$v_{em} = \alpha \rho / \mu_0 d. \quad (45)$$

Here $\alpha \simeq 1$ is a numerical factor depending on the sample geometry and type of mode, and $\mu_0$ is the vacuum magnetic permeability.

As discussed in the previous section, Eq. (45) properly describes the propagation velocity of the dendrite's trunk, which is heated to a temperature close to $T_c$. Coating by a normal film decreases the local resistivity, and therefore, decreases the trunk velocity. This is the physical reason for the refraction of avalanche branches.

The quantitative estimates are as follows.[39] For a superconducting film of thickness $d_s$ and resistivity $\rho_s$, coated by a metal layer of thickness $d_m$ and resistivity $\rho_m$, one can define an effective resistivity $\rho_c$. If there is no exchange of electrical charge between the two layers, the resistivity of the coated film is given by

$$\rho_c = (d_s + d_m)\left(\frac{d_s}{\rho_s} + \frac{d_m}{\rho_m}\right)^{-1}. \quad (46)$$

From Eq. (45) it then follows that the propagation velocity in the bare superconducting film, $v_s$, and the velocity in the coated film, $v_c$, are related by

$$\frac{v_s}{v_c} = 1 + \frac{\rho_s d_m}{\rho_m d_s}. \quad (47)$$

Thus, from Snell's law, the relative refractive index for rays propagation between coated and bare areas of a superconducting film is given by the rhs of Eq. (47). The ratio $(\rho_s d_m)/(\rho_m d_s) \equiv S$ was introduced recently[64] as a parameter to quantify how efficiently a metal coating will suppress flux avalanches in an adjacent superconductor. Using again $n = 1.38$, we find for the present system that $S = 0.38$. Compared with the case considered in Ref. 64, where $S \gg 1$ and the metal coating caused rapid decay of the avalanches, the present $S$-value represents weak damping, which evidently is a prerequisite for refraction of the branches to be observed.

With the values for $d_s$ and $d_m$ in the present sample, one finds $\rho_s \approx 0.07 \rho_m$. From this it follows that the instantaneous temperature at the front of a propagating avalanche is not far from the superconductor's critical temperature. Also this is consistent with the assumption that the front propagation can be considered analogous to that of the modes introduced in Refs. 82 and 83.

To visualize the refraction taking place at the lower edge of the strip, we show in Fig. 13, lower panel, a set of straight dotted lines drawn parallel to the refracted rays in the bare superconductor region above the strip. The construction presumes that Snell's law with same index of refraction applies also at the lower edge, and it turns out that all lines meet in one point. This strongly suggests that the rays originate from one single event at an intermediate stage of the avalanche. In the same panel one can make another interesting observation, namely a clearly visible example of dendrite reflection. The event takes place at the lower edge of the strip, and the reflected ray is drawn as a dashed line at an angle equal to that of the incident ray.

## 8. Anisotropic avalanche activity

### 8.1. Fixed anisotropy

In 2007 a remarkable observation was reported by Albrecht et al.,[69] who presented MO images of a $5 \times 5$ mm film of $MgB_2$ deposited on a vicinal $Al_2O_3$ substrate. Due to the slight tilt relative to a main crystallographic axis the substrate surface had an array of linear steps of one unit cell in height and separated by 27 nm. The steps were aligned approximately along one pair of the film edges. Above 10 K the sample was thermomagnetically stable, and only regular gradual penetration of flux was observed as the applied perpendicular magnetic field increased. The images revealed also that the pinning of vortices moving perpendicular to the surface steps was larger than for the vortices moving parallel to the steps. In terms of critical current density, it was found quantitatively that $J_c^L/J_c^T = 1.06$, where $J_c^L$ and $J_c^T$ are the critical densities of currents flowing along and transversely to the steps, respectively.

Although small, this 6% anisotropy had a dramatic impact on the flux penetration below 10 K, the threshold temperature below which this $MgB_2$ film became thermo-magnetically unstable. Well below 10 K the avalanches nucleated evenly from all 4 edges of the sample, see Fig. 14 (upper). However, close to 10 K, the lower image reveals that they occurred only from the pair of edges where the larger critical current was flowing.

This striking behavior was explained based on theoretical results obtained earlier in works by Denisov et al.[57,58] Within their model, a film of thickness, $d$, becomes unstable when the flux penetration front reaches a depth, $\ell_x$, given by

$$\ell_x = \frac{\pi}{2}\sqrt{\frac{\kappa T^* d}{J_c E}}\left(1 - \sqrt{\frac{2h_0 T^*}{nJ_c E}}\right)^{-1}. \quad (48)$$

The threshold value for the applied perpendicular field, $H_{th}$, can then be found by combining Eq. (48) with the Bean model expression for the flux penetration depth in a thin strip of width $2w$,[85,86] which gives

$$H_{th} = \frac{J_c}{\pi}\operatorname{acosh}\left(\frac{1}{1 - \ell_x/w}\right). \quad (49)$$

Shown in Fig. 15 as a full curve is the relation between the threshold field and the critical sheet current. The graph is based on the two equations above using material parameters representing a film of $MgB_2$, i.e., $kT^*/E = 140$ A and $h_0 T^*/nE = 9230$ A/m, which can mean, e.g., $T^* = 10$ K, $E = 0.01$ V/m, $\kappa = 0.14$ W/K m, $n = 30$, and $h_0 = 280$ W/(K m)$^2$.[69]

Included in the plot are also 3 pairs of vertical lines representing two critical sheet currents differing only slightly in magnitude. The lines are drawn vertical, consistent with the Bean model approximation. At low temperatures, the full curve is nearly horizontal, i.e., the threshold field $H_{th}$ is essentially independent of $J_c$. This corresponds to what was observed at 8 K in the $MgB_2$ film. At increasing temperatures, both $J_c$'s are reduced, and when approaching 10 K the graph shows that the pair of threshold fields separate by increasing amounts. It follows from the graph that near 10 K the avalanche activity will start first from the edges where the largest critical current flows, which is exactly what the MOI observations revealed. Then, at even higher temperatures the two $J_c$'s are reduced further, and in the graph they both eventually enter the range where the theory predicts stable flux penetration behavior, again in full accord with the experiments in Ref. 69.

Evidently, when anisotropic flux dynamics in a superconducting film is a consequence of the substrate's surface structure, the anisotropy can hardly be changed or manipulated after the film has been synthesized. However, quite recently, a different approach was found which allows to reversibly change and control the anisotropy in the flux dynamics of superconducting films.

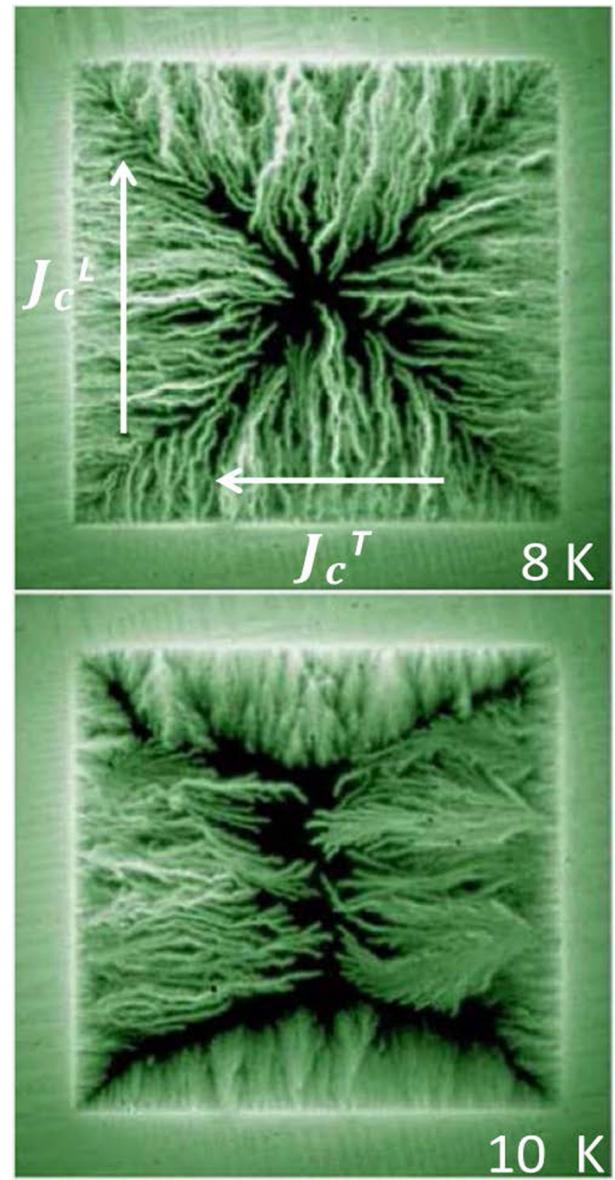

Fig. 14. Magneto-optical images of flux penetration in a 200 nm thick $MgB_2$ film grown on a vicinal substrate. The steps in the substrate are aligned approximately vertical in the figure. The upper and lower images were recorded at 8 and 10 K, respectively. The non-uniformity in the penetration at 10 K from the two horizontal edges is due to edge roughness and other minor sample imperfections. Adapted from Ref. 69.

### 8.2. Tunable anisotropy

In 2016 Vlasko-Vlasov et al.[87] reported MOI studies of Nb films deposited by magnetron sputtering on Si(100) substrates. Films of two thicknesses, 100 and 200 nm, and $T_c$ near 9 K were shaped as squares with sides 2.0 and 2.5 mm, respectively. When cooled in the presence of an in-plane magnetic field the thicker film, when it subsequently was exposed to an increasing perpendicular field, displayed large anisotropy in the flux penetration pattern. When the same procedure was applied to the thinner film, it showed essentially isotropic flux penetration. This qualitative difference in behavior was attributed to the presence of frozen-in in-plane vortices in the thicker film, while the thinner film was too thin to accommodate in-plane vortices.

Shown in Fig. 16, left panel, is an example of anisotropic flux penetration in a 200 nm thick Nb film, where the indicated in-plane field $H_{\parallel} = 1$ kOe was applied during the cooling to 7 K. The image was recorded after adding a perpendicular field of $H_\perp = 20$ Oe. Quantitative measurements[89] of the anisotropy in the critical sheet currents, $J_c^T$ and $J_c^L$, see Fig. 16, right panel, found that their ratio is well described by the qubic dependence,[88]

$$J_c^L/J_c^T = 1 + cH_\parallel^3,$$

with $c = 8 \times 10^{-10}$ Oe$^{-3}$.

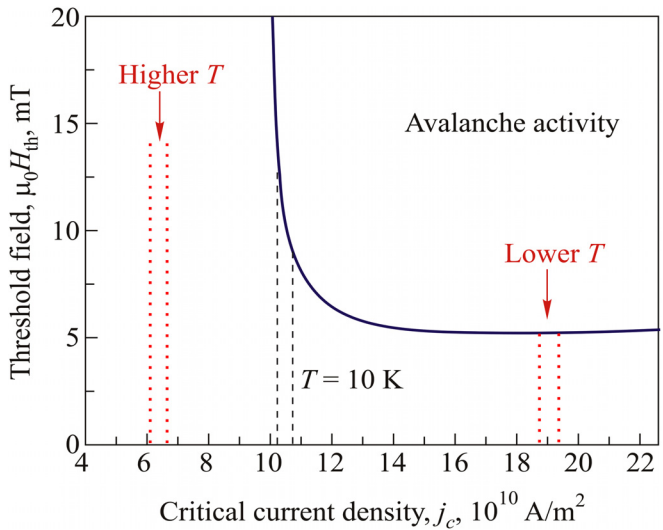

Fig. 15. Graph of the threshold perpendicular magnetic field versus critical current density, for onset of avalanche activity in films of MgB$_2$ (full curve). The pairs of dashed/dotted lines show the critical current density at 3 temperatures, and the two lines in each pair indicate the anisotropy in $j_c$.

Separate measurements were required to decide whether the anisotropy is due to reduced pinning of the perpendicular vortices when moving parallel the frozen-in in-plane vortices, or enhanced pinning of perpendicular vortices traversing the array of the in-plane ones, or both. To resolve this question a local flux injector,[89] was used, where the square Nb sample was extended by two strips forming an inverted V-shape allowing for a transport current to be passed through a small region of the square near its lower edge, see Fig. 17.

Shown in the left panel is an image of the flux penetration caused by passing a current pulse of 0.6 A after the film had been initially zero-field cooled to 7 K. The current pulse lead to penetration of flux in an area with shape close to a semi-circle. When applying the same pulse after the film was cooled in the presence of $H_\parallel = 1$ kOe aligned as indicated in the figure, the area of injected flux was distorted by a significant elongation in the direction aligned with the frozen-in flux. Moreover, one sees that the horizontal width of the area is essentially the same as that in panel (a). This shows that freezing in the field $H_\parallel$ leaves $J_c^L$ essentially unchanged, whereas $J_c^T$ becomes smaller.

Striking consequences of this effect was found in the flux dynamics at lower temperatures, where the penetration of perpendicular flux is dominated by avalanche activity. Presented in Fig. 18 are images of the flux penetration in a plain square Nb film, similar to that displayed in Fig. 16. In Fig. 18 panels (a)–(d) the film was initially cooled to 2.5 K in the presence of in-plane fields of magnitudes, 0, 0.7, 1.0 and 1.5 kOe, respectively. Then, a perpendicular field of $H_\perp = 38$ Oe was applied, triggering dendritic avalanches, which are seen to dominate the flux penetration in all four panels. Each dendritic structure is the result of one avalanche event, and is not seen to change thereafter. All the avalanches start from separate nucleation points along the edge.

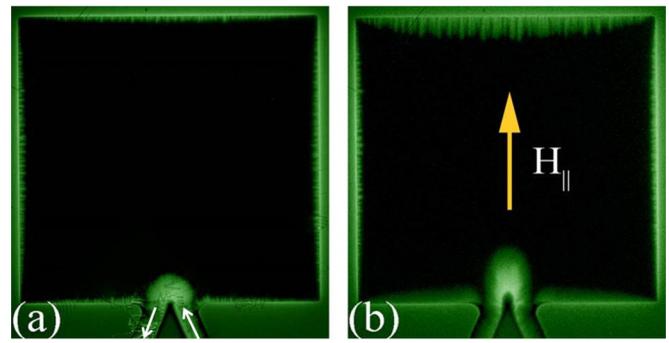

Fig. 17. Magneto-optical images of the Nb film at 7 K after a current pulse (white arrows) was passed through a pair of strips extending the sample by an inverted V-shape at the lower edge. In (a) the film was initially zero-field cooled, and in (b) it was cooled in the presence of an in-plane field of $H_\parallel = 1$ kOe.

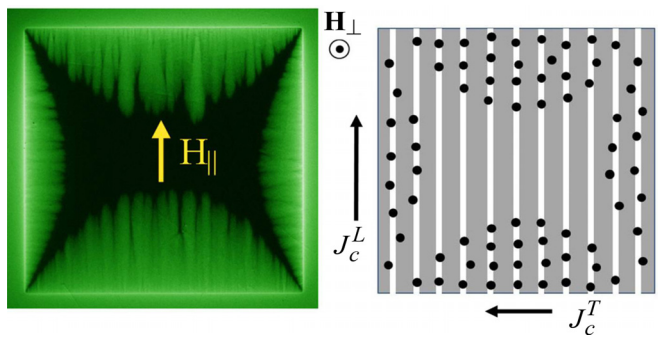

Fig. 16. Left: Magneto-optical image showing field-induced anisotropic flux penetration in a 2.5 × 2.5 mm Nb film of thickness 200 nm. The in-plane field $H_\parallel = 1$ kOe was frozen in during the initial cooling to 7 K. From Ref. 88. Right: Illustration of anisotropic penetration of perpendicular vortices (black dots) in the presence of frozen-in in-plane vortices (white lines). From Ref. 88.

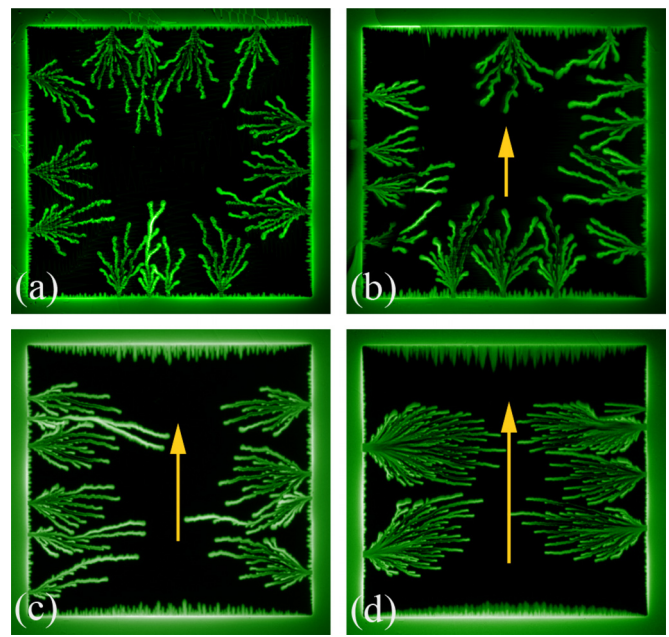

Fig. 18. Magneto-optical images of the penetration of perpendicular flux in a square Nb film, where in-plane fields, indicated by the arrows, were applied during the initial cooling to 2.5 K. In panels (a)–(d) the $H_\parallel$ were 0, 0.7, 1.0 and 1.5 kOe, respectively. From Ref. 88.

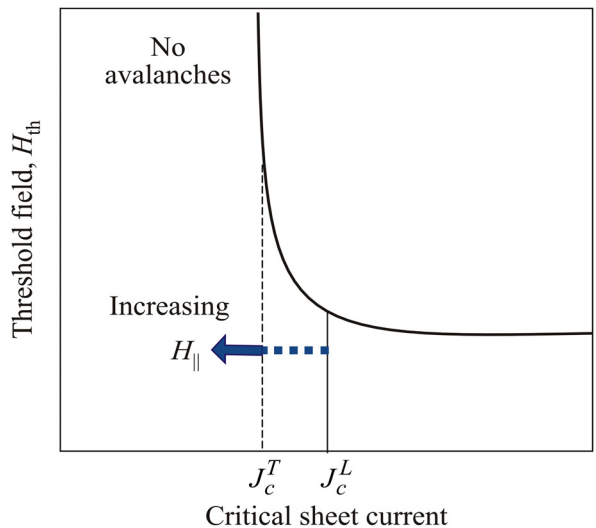

Fig. 19. Generic curve for threshold applied perpendicular field for the onset of thermomagnetic avalanche activity in superconducting films versus their critical sheet current. From Ref. 88.

In panel (a) of Fig. 18 one sees that the avalanches nucleated from locations quite evenly distributed between all 4 edges. However, in panel (b) the isotropic symmetry is broken as the majority of avalanches here nucleate from the pair of edges that are aligned with the frozen-in field, $H_\parallel$ Then in panel (c), the anisotropy is complete, as no avalanche nucleated from the edges perpendicular to $H_\parallel$. When increasing the $H_\parallel$ further, the full anisotropy remains, and the avalanches become fewer but larger in size, see panel (d).

Also much of this behavior can be explained from Eqs. (48) and (49), and the generic graph of the threshold magnetic field versus critical sheet current, see Fig. 19 In this plot the full vertical line represents $J_c^L$, the critical sheet current flowing parallel to the frozen-in vortices, see Fig. 16 (right). As found experimentally, this line remains fixed in the graph, being essentially independent of $H_\parallel$.

The dashed line, representing $J_c^L$, should for the isotropic case, $H_\parallel = 0$, obviously overlap with $J_c^T$. Then, as $H_\parallel$ increases, the $J_c^T$ is gradually reduced, and the dashed line shifts to the left in the graph. The threshold field increases for avalanche nucleation along the edges where $J_c^T$ flows. At the same time, the threshold field at the other pair of edges remain unchanged. Thus, more avalanche events are expected to start there, in full accord with the anisotropy seen in Fig. 16(b).

As $H_\parallel$ increases even further, the dashed line in Fig. 19 at some point will enter the region where avalanches can no longer occur. Thus, avalanches will then only nucleate from the two edges along which the $J_c^L$ flows, again in full agreement with the MOI observations. The entire scenario of different avalanche activities is therefore qualitatively explained.

Note here also the similarity in the flux avalanche patterns in Fig. 14 (upper) and (lower), and in Figs. 18(a) and 18(d), respectively. The two quite different systems display the same change in the avalanche behavior in spite that the origin of anisotropy is quite different in these two cases.

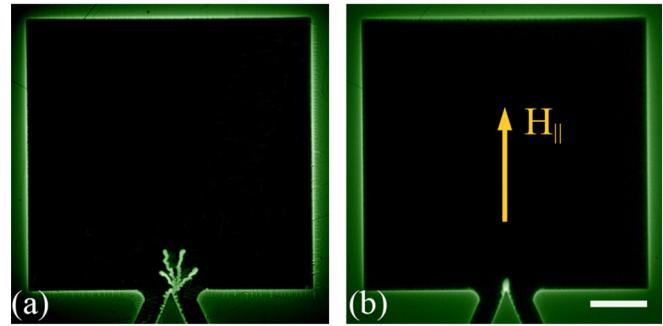

Fig. 20. Magneto-optical images of the penetration of perpendicular flux in a square Nb film extended with an inverted V-shaped flux injector at the lower edge. In panel (a) the sample was initially cooled in zero magnetic field, and in (b) it was cooled while applying an in-plane field $H_\parallel = 1.5$ kOe oriented as indicated by the arrow. Both panels show the flux distribution after a current pulse of 1 A was passed through the injector. The scale bar is 0.5 mm long.

### 8.3. Active triggering of avalanches

When the inverted V-shaped flux injector is activated by passing a current puls at a sufficiently low temperature, the result can be to trigger an avalanche event. Shown in Fig. 20(a) is an example of an avalanche triggered by a pulse of magnitude 1.0 A and duration 200 ms. The 200 nm thick Nb film was here initially zero-field cooled to $T = 2.5$ K. As expected, the avalanche was rooted at the flux injection point, and displayed a dendritic morphology, which when repeating the experiment never reproduced itself.

Interestingly, when the flux injection experiment was carried out when the same sample was initially cooled in the presence of an in-plane field of $H_\parallel = 1.5$ kOe oriented as shown in panel (b) of Fig. 20, the behavior changed dramatically. This image shows that in this field-cooled condition the avalanche was not allowed to develop beyond its incipient stage.

Again, this can be explained by the fact that a frozen-in field $H_\parallel$ shifts $J_c$ flowing in the transverse direction to smaller values. With the $H_\parallel$ frozen-in as indicated in Fig. 20, the $J_c^T$—the current density flowing along the edge where the injector is located, becomes too small for a finite threshold field to exist. Thus, the $H$-induced reduction in $J_c^T$ stabilizes the superconducting film with respect to onset of avalanche activity.

### 9. Conclusions

In this paper, we have reviewed recent theoretical and experimental work on thermomagnetic instability leading to magnetic avalanches in thin-film superconductors. Our theory is macroscopic—it is based on analysis of the Maxwell equations and local thermal balance between the Joule heat release and its spreading along the film and into the substrate. The properties of the material are taken into account through realistic nonlinear current-voltage curve, as well as through the thermal characteristics of the system.

Starting from the magnetic flux distribution in the critical state we first performed the linear stability analysis. That was done analytically, and as a result explicit onset conditions, i.e., thresholds in temperature, electric field and applied magnetic field were obtained as functions of material parameters. We considered both bare films and the films coated by a layer of a normal metal allowing to control the stability regime.

The resulting stability diagram demonstrates a rich physical picture showing several regimes of the thermomagnetic instability including both monotonous and oscillatory growing modes. The oscillatory modes are more unstable than the monotonous ones. As a result, large-scale avalanches can nucleate directly from the Bean critical state, rather than being mediated by non-thermal microavalanches, which up to now was the most plausible explanation for the occurrence of dendritic avalanches in films during slow field variations.

The analytical work is supplemented by numerical simulations allowing to analyze the propagation of dendritic avalanches at different stages. As a result of the analysis characteristic time scales for the thermomagnetic instability were revealed. In particular, the striking phenomenon of ultra-fast propagation of the avalanches is now understood. We present main concepts of the numerical procedure we have used.

In the rest of the paper we analyzed several manifestations of the thermomagnetic instability observed experimentally using magneto-optical imaging. This method turned out to be extremely fruitful since it possesses both sufficiently high spatial and temporal resolution. As an example of specific features of the instability we discuss the experimentally observed ray-optics behavior of the dendrites' trunks. To observe such a behavior samples coated by strips of normal metal were used. Another example is observed dramatic anisotropy of the flux patterns observed in weakly anisotropic samples. We present main experimental results regarding the aforementioned phenomena and provide the explanations based on the theory described in the first part of the paper.

To summarize, we conclude that main observed features of the thermomagnetic instability in thin superconducting films are now understood.

T.H.J. acknowledges the hospitality of Professor Barbara Neuhauser during his sabbatical stay at San Francisco State University.

a)Email: j.i.vestgarden@smn.uio.no
b)Email: t.h.johansen@fys.uio.no
c)Email: iouri.galperine@fys.uio.no